\documentstyle[aps,prb,,multicol,psfig,amssymb]{revtex}
\newif\ifmynarrow 
\mynarrowfalse
\renewcommand{\narrowtext}{%
  \ifmynarrow\hspace*{\fill}\raisebox{-1ex}[0pt][0pt]{%
    \rule{0.3pt}{1ex}%
    \rule[1ex]{20.5pc}{0.3pt}}\fi
  \mynarrowtrue
  \vspace{-1.0ex}%
  \begin{multicols}{2}%
  \par\global\columnwidth20.5pc
  \global\hsize\columnwidth\global\linewidth\columnwidth
  \global\displaywidth\columnwidth}
\renewcommand{\widetext}{%
  \end{multicols}%
  \vspace{-2.5ex}%
  \noindent\raisebox{1ex}[0pt][0pt]{%
    \rule{20.5pc}{0.3pt}%
    \rule{0.3pt}{1ex}}%
  \par\global\columnwidth42.5pc
  \global\hsize\columnwidth\global\linewidth\columnwidth
  \global\displaywidth\columnwidth}
\begin{document}
\draft 
\tighten 
\title{Manipulation of spin dephasing in InAs quantum wires} 
\author{J. L. Cheng$^{1,2}$, M. Q. Weng$^{2}$, and M. W. Wu$^{1,2}$
\footnote{Email: mwwu@ustc.edu.cn}}
\address{Structure Research Laboratory, University of Science \&%
  Technology of China, Academia Sinica, Hefei, Anhui, 230026, China}
\address{Department of Physics, University of Science \&%
  Technology of China, Hefei, Anhui, 230026, China
\footnote{Mailing address.}
}
\date{\today}
\maketitle
\begin{abstract}
  The spin dephasing due to the Rashba spin-orbit coupling, especially 
its dependence on the direction of the electric field
is  studied in InAs quantum wire. 
We find that the spin dephasing is strongly affected by the
  angle of Rashba effective magnetic field and the applied magnetic
  field. The nonlinearity in spin dephasing time versus the direction
of the electric field shows a
  potential revenue to manipulate the spin lifetime in spintronic
  device. Moreover, we figure out a quantity that can well represent
  the inhomogeneous broadening of the system which may help us to
  understand the many-body spin dephasing due to the Rashba effect.

\end{abstract}

\narrowtext

Since Datta and Das proposed a new type of electronic transistor that
utilizes the electron spin freedom,\cite{datta_1990} many 
efforts have been devoted to the realization of the spintronic 
devices.\cite{spintronics,wolf_sci_2001,spintronics_awsch} As the
function of these devices rely on the spin coherence, the key to
realize these devices is to manipulate the spin coherence.  The
confined narrow band gap semiconductor systems such as quantum well
and quantum wire 
are proposed to be good candidates for spintronic devices as the
electron spin precession in these systems can be easily tuned by
external gate voltages through the Rashba effect\cite{rashba} which states
that the spin-orbit coupling is proportional to the applied and/or interface
electric field.  In quantum wells, the direction of electric field
(DOEF) is
always perpendicular to the interface. Therefore the manipulation of the
spin coherence can only be realized through the magnitude of the
electric field. While in quantum wire, both the direction and the
magnitude of the electric field can be tuned by the external gate
voltages.  This  provides one more degree of freedom to manipulate the spin
coherence by alerting the DOEF.
Here we demonstrate the possibility 
of using this additional degree of freedom to realize the manipulation of spin
coherence in $n$-typed InAs quantum wire by calculating the dependence
of the spin dephasing time (SDT) with the DOEF
for different spin configurations.

It is known that for high temperature the spin dephasing
magnetism for low dimensional $n$-typed
InAs is due to the Rashba effect.\cite{rashba} We note
here that the Rashba effect provides two spin dephasing channels:
The first one is the effective spin-flip (SF) scattering caused by
the anisotropy of the Rashba term and the spin conserving (SC)
scattering. This dephasing effect has been widely discussed in the literature.
The second one is a newly proposed many-body effect which is caused by
the intrinsic inhomogeneous broadening ({\em i.e.} $k$-dependence)
from the Rashba term itself together with 
the SC scattering. Our previous works have shown that this many-body
effect plays important, sometimes dominant, role in spin
dephasing\cite{wu_epjb_2000,wu_js_2001,c0302330,c0210313,c0303169} as
well as in spin transport.\cite{weng_prb_2002,weng_jap_2003} In this
paper, we apply the many-body kinetic theory developed in our
previous papers\cite{wu_pss_2000,c0302330,c0210313,c0303169} to study the
spin dephasing in InAs quantum wire (QW) and show the
feasibility of the manipulation of the spin coherence
through changing the DOEF.
Moreover, through changing the DOEF
we are able to change the Rashba term and thus the
inhomogeneous broadening {\em alone} while keeping the SC scattering
unchanged. In this paper, we are going to have the inhomogeneous
broadening quantified for the first time and
show how the many-body spin
dephasing depends on the inhomogeneous broadening.

\begin{figure}[htb]
\psfig{figure=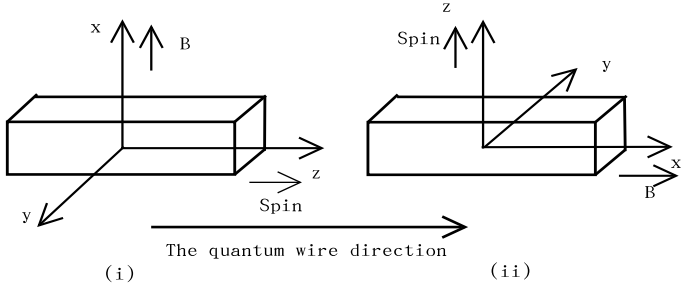,width=9.cm,height=4cm,angle=0} 
\vskip 5pt
\caption{The two special configurations we study. 
  (i) the axis of the QW is laid along the $z$ direction; 
  (ii) the axis of the QW is laid along the $x$ direction.
}
\label{fig:config}
\end{figure}

The QW system we study is an InAs nano-structure confined by
square potential wells with width $a$ in both perpendicular
directions. In this system, the electron state is
characterized by two subband indexes $n_1$ and $n_2$ which stand for the
quantum number of two confined direction respectively, the 1D momentum
${\bf k}$
along the wire together with a spin index $\sigma$. 
In the present paper, we only consider the electron density  that the
subband separation is 
large enough so that only the lowest subband is populated and the
transition to the upper subband is negligible. Therefore, we need
only consider the subband with
$n_1=n_2=1$, and drop the subband index consequently. 
In this study, the spin quantization direction is chosen to be
$z$-axis. A moderate magnetic field is applied along the $x$-axis. 
Taking account of the Rashba effect, the electrons experience 
an additional wavevector dependent effective magnetic field 
\begin{equation}
  \label{eq:Rashba}
  {\bf h}({\bf k}) = \alpha_0 {\bf E}\times{\bf k}\,,
\end{equation}
where $\alpha_0$ is the Rashba spin-orbit parameter and ${\bf E}$ is the 
interface electric field. It is note that this electric field
can be tuned by the gate voltages on both confined
directions. Therefore, both the magnitude and the direction of this
electric field can be tuned in the plane perpendicular to the wire.
Under the two special configurations (see
Fig.\ref{fig:config}) we study, the
electric field and the Rashba effective magnetic field (REMF)
are: (i)
${\bf E}=(E \sin\theta, E \cos\theta, 0)$, 
${\bf h}(k) = \alpha E k(\cos\theta, -\sin\theta, 0)$
for the first configuration where the axis of QW is
laid along $z$-direction; and (ii) 
${\bf E}=(0, E\cos\theta, E\sin\theta)$, 
${\bf h}(k)=\alpha k E (0, -\sin\theta, \cos\theta)$
for the second configuration where the axis of QW is
laid along the $x$-direction respectively. Here $\theta$ is the angle
between the electric field and the $y$-axis. 
In configuration (i) it is also the angle
between the REMF and the AMF.
The interaction Hamiltonian $H_I$ is composed of Coulomb interaction
$H_{ee}$, electron-phonon interaction $H_{ph}$, as well as
electron-impurity scattering $H_i$. Their expressions can be found in
textbooks.\cite{haug,mahan} 

We construct the kinetic Bloch equations by the nonequilibrium Green
function method\cite{haug} as follows:
\begin{equation}
  \label{eq:bloch}
  \dot{\rho}_{{\bf k},\sigma\sigma^{\prime}}
  =\dot{\rho}_{{\bf k},\sigma\sigma^{\prime}}|_{\mbox{coh}}
  +\dot{\rho}_{{\bf k},\sigma\sigma^{\prime}}|_{\mbox{scatt}}\ .
\end{equation}
Here $\rho_{{\bf k}}$ represents the single particle density
matrix. The diagonal elements describe the electron distribution
functions $\rho_{{\bf k},\sigma\sigma}=f_{{\bf k}\sigma}$. 
While the off-diagonal elements 
$\rho_{{\bf k},{1\over 2}-{1\over2}}\equiv\rho_{{\bf k}}$ 
are the inter-spin-band polarizations
(coherence) of the spin coherence.\cite{wu_prb_2000} Note that 
$\rho_{{\bf k},-{1\over 2}{1\over 2}}\equiv 
\rho^{\ast}_{{\bf k},{1\over 2}-{1\over 2}}=\rho^{\ast}_{{\bf k}}$. 

The coherent part of the kinetic equations for the electron
distribution function and the spin coherence are 
\begin{equation}
  \label{cohf}
  \dot{f}_{k,\sigma}|_{coh} = -2\sigma
  \mbox{Im}[\rho_{k}(R_{k,\downarrow\uparrow}-
  \frac{1}{2}g\mu_BB-\sum_qV_q\rho^{\star}_{k-q})], 
\end{equation}
and
\begin{eqnarray}
  \label{cohrho}
  \dot{\rho}_{k}|_{coh} &=&
  -i[(R_{k,\uparrow\uparrow}-
  R_{k,\downarrow\downarrow}-
  \sum_qV_q(f_{k-q,\uparrow}-f_{k-q,\downarrow}))\rho_k
  \nonumber\\ &+&
  (R_{k,\uparrow\downarrow}-\frac{1}{2}g\mu_BB-
  \sum_qV_q\rho_{k-q}(f_{k\uparrow}-f_{k\downarrow})]
\end{eqnarray}
In these equations, the terms with $R_{k,\sigma\sigma^{\prime}}$ come
from the Rashba term, and 
$R_{k,\sigma\sigma^{\prime}}=-\alpha_0
Ek\delta_{\sigma-\sigma^{\prime}}\{\sin\theta+i\sigma\cos\theta\}$
for configuration (i) and 
$R_{k,\sigma\sigma^{\prime}}=\alpha_0
Ek\{\delta_{\sigma\sigma^{\prime}}\cos\theta+
i\delta_{\sigma-\sigma^{\prime}}\sin\theta\}$ 
for configuration (ii). 
While the term with $V_q$ is the contribution of the electron-electron
interaction up to HF approximation with $V_q$ denotes the 1D Coulomb
matrix element under static screening. 
The scattering term ${\partial f_{k\sigma}\over \partial
  t}|_{\mbox{scatt}}$ and ${\partial \rho_{k}\over \partial
  t}|_{\mbox{scatt}}$ are listed detail in
Refs.~\onlinecite{c0302330,c0210313,c0303169}, and will not be
repeated here.

The initial conditions at $t=0$ are taken as: 
\begin{eqnarray}
  &&\rho_{\bf k}|_{\rm t=0} = 0
  \label{eq:rho_init}\\
&&  f_{{\bf k}\sigma}|_{\rm t=0} = 1/\bigl\{\exp[(\varepsilon_{\bf
    k}-\mu_{\sigma})/k_BT]+1\bigr\} 
  \label{eq:fk_init}
\end{eqnarray}
where $\mu_\sigma$ is the chemical potential for spin $\sigma$. The
condition $\mu_{\frac{1}{2}}\neq\mu_{-\frac{1}{2}}$ gives rise to the
imbalance of the electron densities of the two spin
bands. Eqs.~(\ref{eq:bloch}) through (\ref{cohrho}) together
with the initial conditions Eqs.~(\ref{eq:rho_init}) and
(\ref{eq:fk_init}) comprise the complete 
set of kinetic Bloch equations of our investigation.

\begin{figure}[htb]
\psfig{figure=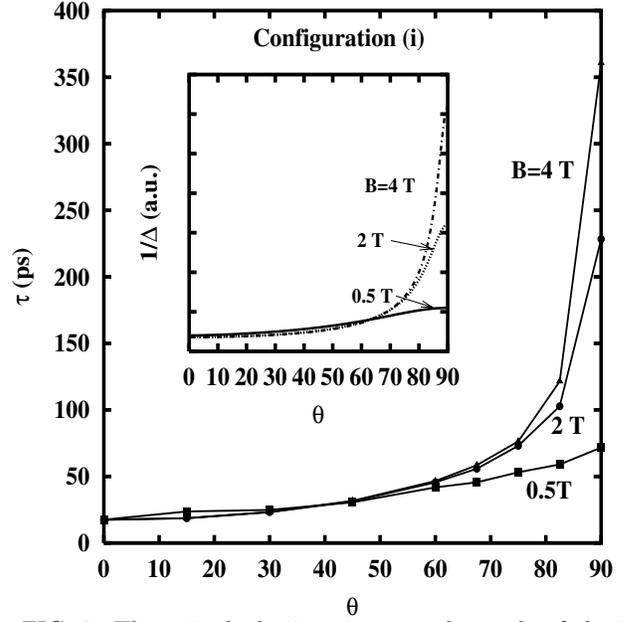,width=8.5cm,height=8.5cm,angle=0}
\caption{
  The spin dephasing time vs. the angle of the interface electric
  field and $y$-axis for configuration (i) shown in
  Fig.~\ref{fig:config} under different magnetic field:
  ($\blacksquare$) B=0.5~T; ($\bullet$) B=2~T; ($\blacktriangle$) and
  B=4~T. The inset shows the standard deviation of the magnetic field
  the electron surfers. 
}
\label{conf1}
\end{figure}

By numerically solving the kinetic Bloch Eqs (\ref{eq:bloch}) and the
initial condition (\ref{eq:rho_init}) and (\ref{eq:fk_init}), we are
able to get the temporal evolution of the electron distribution
function and the spin coherence self-consistently.  As discussed in
the previous papers,\cite{wu_prb_2000,kuhn} irreversible spin dephasing
can be well defined by the slope of the envelope of the incoherently
summed spin coherence $\rho(t)=\sum_{{\bf k}}|\rho_{{\bf k}}|$.  In
this way, we obtain the SDT which is defined as the
inverse of the decay rate of the envelope of $\rho(t)$. 

We calculate the SDT in an InAs QW under two
configurations shown in Fig.~\ref{fig:config} at temperature
$T=150$~K. With this relative high temperature we 
only need to consider the electron-longitudinal optical phonon, the
electron-impurity and the electron-electron Coulomb scattering.
In the calculation the interface electric field
$E$, the density of the electron $N_e$, the well widths 
of the QW $a$ and the density of the impurities
are chosen to be $1.4\times 10^4$~V/cm,
$5.3\times 10^{16}$~cm$^{-3}$, $15$~nm and 0 respectively. 
The Rashba spin-orbit parameter $\alpha_0$ is
$110\AA^2$\cite{lommer}, 
and the other material parameters are cited from Ref. 
\onlinecite{made}.
By changing the direction of the electron field we are able to change
the direction of the REMF. For configuration (i),  
by changing the DOEF, the angle between  the AMF
and REMF can be changed but both the
AMF and the REMF are perpendicular to the spin
polarization.  While for configuration (ii), when the
direction of the DOEF changes, the REMF is
always normal to the AMF but the 
angle between the REMF and the spin
polarization  changes.  With the different relative
directions between the AMF, the REMF and the spin polarization, one
may expect that the manipulation of the spin dephasing via
the  DOEF can be quite different under
different configurations.
 
\begin{figure}[htb]
\psfig{figure=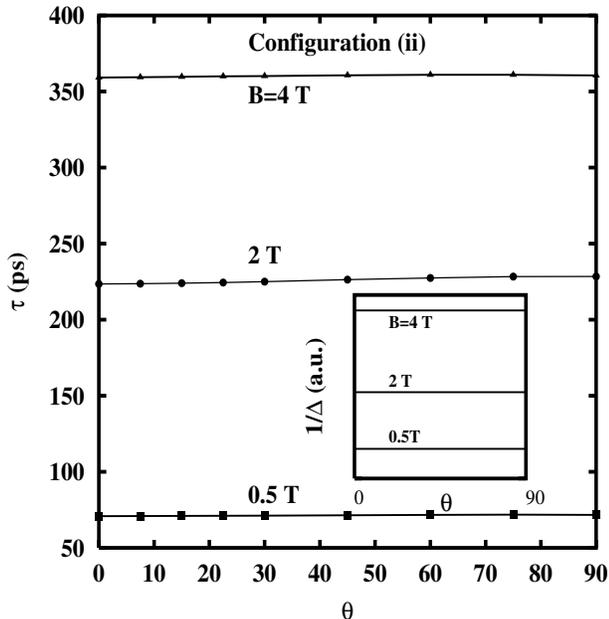,width=8.5cm,height=8.5cm,angle=0}
\caption{
  The spin dephasing time vs. the angle of the interface electric
  field and $y$-axis for configuration (ii) shown in
  Fig.~\ref{fig:config} under different magnetic field:
  ($\blacksquare$) B=0.5~T; ($\bullet$) B=2~T; ($\blacktriangle$) and
  B=4~T. The inset shows the standard deviation of the magnetic field
  the electron surfers. 
}
\label{conf2}
\end{figure}

Our results are plotted as functions of the angle $\theta$
under different AMF's in Figs.~\ref{conf1} and \ref{conf2} 
for the two special configurations. 
One notices that for configuration (i) when the REMF
is parallel to the AMF, 
{\em i.e.} $\theta=0$, the SDT is insensitive to the magnetic
field. Moreover, the SDT is also insensitive to $\theta$ when it is
smaller than $60^{\circ}$. After that, SDT gets a rapid increase as $\theta$
increases for all of the AMF we study. For example
when $B=2$~T, the SDT at $\theta=90^{\circ}$ is 
3 times higher than that at $\theta=60^{\circ}$. 
The larger the AMF, the faster  the SDT  increases with $\theta$. 
As a result, the SDT increases with the AMF when
the REMF is near the normal of the AMF.
However, the result of the configuration (ii) is quite different: 
Here the SDT is independent on the DOEF  under a
moderate AMF. Nevertheless, it is noted that the SDT 
also increases with the AMF and the value is 
equal to the corresponding one in configuration (i) when
the REMF is normal to the spin polarization, {\em ie.}, $\theta=90^{\circ}$.

The DOEF dependence of spin dephasing can be understood from the
many-body point of view. Besides the single particle spin dephasing
due to the effective SF scattering, Rashba effect also provides an
additional many-body spin dephasing channel through its inherit
inhomogeneous broadening. It has been pointed out
before the inhomogeneous broadening together with the SC scattering
provides an irreversible spin dephasing. 
In the present case, the inhomogeneous broadening can be 
calculated as follows: The electrons in QW undergo the
Larmor precession around the
magnetic field composed of the AMF and the REMF.
As the REMF is
wavevector dependent, the frequency and the direction of the
Larmor precession differs between electrons with different
wavevectors. This is the origin of the inhomogeneous
broadening. Therefore, one can expect that the inhomogeneous
broadening of the frequency of Larmor precession may somehow reflect
the spin dephasing due to the inhomogeneous broadening. Starting from
this point, we carry out the corresponding standard deviation $\Delta$
of the Larmor frequencies:
\begin{eqnarray}
  \label{eq:delta}
\Delta^2&=&
  \langle
  (g\mu_BB+\alpha_0E\,k\cos\phi)^2+(\alpha_0E\,k\sin\phi)^2
  \rangle\nonumber\\
  & &\mbox{}-\langle
  \sqrt{(g\mu_BB+\alpha_0E\,k\cos\phi)^2+(\alpha_0E\,k\sin\phi)^2}
  \rangle^2
\end{eqnarray}
with $\phi$ denoting the angle between ${\bf B}$ and ${\bf h}({\bf
  k})$. In configuration (i), $\phi$ is  $\theta$ and in
configuration (ii) $\phi$ is always $90^{\circ}$. 
$\langle \cdots \rangle$ represents the average over the
imbalance of the spin-up and spin-down electrons: 
\begin{equation}
  \langle A(k)\rangle 
  ={\int dk\;(f_{k\,\uparrow}-f_{k\,\downarrow}) A(k)\over
    \int dk\;(f_{k\,\uparrow}-f_{k\,\downarrow})}\;. 
\end{equation}

In the inset of  Figs.~\ref{conf1} and \ref{conf2},  $1/\Delta$
is plotted as a function of the $\phi$ ($\theta$) 
for two configurations.  
One can easily see that the $1/\Delta$-$\theta$ curve is much similar
to the $\tau$-$\theta$ curve: For configuration (i), $1/\Delta$ is 
insensitive to the DOEF and the AMF when
$\theta<60^{\circ}$. When $\theta>60^{\circ}$,  $1/\Delta$
increases rapidly with both the AMF and $\theta$; 
For configuration (ii), $1/\Delta$ does not depend on DOEF,
and its value equals the corresponding one in configuration (i)
with the REMF being perpendicular to the AMF.

From the correspondence between the $\tau$-$\theta$ curves and
the $1/\Delta$-$\theta$ curves one can conclude that the spin dephasing in
the QW we study is determined by the inhomogeneous broadening. With the
$1/\Delta$-$\theta$ curve, it is easy to understand the AMF
dependence and the DOEF dependence of the spin
dephasing: For  configuration (i), when the REMF
is parallel to the AMF, {\em i.e.}
$\theta=0$, $\Delta$ does not depend on the AMF.
Therefore the AMF does not reduce the inhomogeneous
broadening. As a consequence, the spin dephasing is insensitive to the
magnetic field.  Nonetheless when the REMF
is perpendicular to the AMF, that is
$\theta=90^{\circ}$,  $\Delta$ reduces with the AMF.
Consequently, the inhomogeneous broadening, thus the spin
dephasing, is reduced by the AMF.  For 
configuration (ii), the REMF is
perpendicular to the AMF for all directions of the electric field.
Therefore $\Delta$,
the inhomogeneous broadening of the Larmor precession frequency,
is independent of the DOEF and its value is the same to the
corresponding one in  
configuration (i) when the $\theta=90^{\circ}$. As a result, the SDT
in configuration (ii) does not change with $\theta$ and its value
 equals the corresponding
value in  configuration (i) when $\theta$ is $90^{\circ}$. 

The nonlinear property in $\tau$-$\theta$ curve of configuration (i)
can also be understood from the inhomogeneous broadening: 
When the AMF is much larger than the REMF, $\Delta$ is expected to be
\begin{eqnarray}
  &&  \Delta^2=\langle (\alpha_0E\,k)^2\rangle \cos^2\theta
  + {1\over 4(g\mu_BB)^2}\bigl\{
  \langle(\alpha_0E\,k)^4\rangle \nonumber\\
  &&\times  [1-6\cos^2\theta+5\cos^4\theta]
  -\langle(\alpha_0E\,k)^2\rangle^2\sin^4\theta\bigr\}. 
\end{eqnarray}
As the second term is very small, $1/\Delta$ changes with $\theta$
according to $1/\cos\theta$ except in the regime where $\theta$ is
very close to $90^{\circ}$. Consequently, the spin dephasing changes
slowly with $\theta$ when $\theta$ is smaller than $60^{\circ}$ and
then gets a rapid increase after $60^{\circ}$. 

In conclusion we have performed a kinetic study of the spin dephasing in InAs
QW. We find that under a moderate magnetic field, the SDT in some
configuration depends on the DOEF; while in the other configuration,
it does not change with the DOEF. Therefore, for the right
configuration, the spin dephasing can be tuned through changing the
direction of the interface electric field in addition to the magnitude
of the electric field. Moreover, we define a quality that well
represents spin dephasing due to the inhomogeneous broadening of the Rashba
term and show that the spin dephasing in the system we study is
determined  by the inhomogeneous broadening.


\acknowledgments
MWW is supported by the  ``100 Person Project'' of Chinese Academy of
Sciences and Natural Science Foundation of China under Grant
No. 10247002.

\end {multicols}
\end {document}